\documentclass[10pt]{ijnam}
\def\currentvolume{0}
\def\currentissue{0}
\def\currentyear{2017}
\def\month{March}
\pagespan{1}{18}
\copyrightinfo{2017}{} 

\usepackage{
    listings,
    graphicx,
    xcolor,
    enumitem
}
\definecolor{darkgreen}{rgb}{0,0.5,0}

\newcommand*{\Comment}[1]{\hfill #1 }
\newcommand{\pd}[2]{\frac{\partial#1}{\partial#2}} 
\newcommand{\unit}[1]{\ensuremath{\, \mathrm{#1}}} 

\theoremstyle{definition}
\newtheorem{example}{Example}[section]

\newcommand\xqed[1]{\leavevmode\unskip\penalty9999 \hbox{}\nobreak\hfill \quad\hbox{#1}}
\newcommand{\exampleSymbol}{\xqed{$\triangle$}}

\begin{document}

\title[Abstract Elementary Algebra]
{Sparse Automatic Differentiation for Complex Networks of Differential-Algebraic Equations Using Abstract Elementary Algebra}

\author[S. Pele\v{s}]{Slaven Pele\v{s}}
\address{
  United Technologies Research Center, East Hartford, CT 06108\\
  (Present address: Lawrence Livermore National Laboratory, Livermore, CA 94550 USA)
}
\email{peles2@llnl.gov}

\author[S. Klus]{Stefan Klus}
\address{
  United Technologies Research Center, East Hartford, CT 06108\\
  (Present address: Freie Universit\"at Berlin, Department of Mathematics \& Computer Science, Arnimallee 6, 14195 Berlin, Germany.)
}
\email{stefan.klus@fu-berlin.de}



\date{March 27, 2017}



\abstract{Most numerical solvers and libraries nowadays are implemented to use mathematical 
models created with language-specific built-in data types (e.g.\ \texttt{real} in Fortran or 
\texttt{double} in C) and their respective elementary algebra implementations. However, the 
built-in elementary algebra typically has limited functionality and often restricts the flexibility 
of mathematical models and the analysis types that can be applied to those models. 
To overcome this limitation, a number of domain-specific languages such as gPROMS or 
Modelica with more feature-rich built-in data types have been proposed. In this paper, we argue 
that if numerical libraries and solvers are designed to use abstract elementary 
algebra rather than the language-specific built-in algebra, modern mainstream languages 
can be as effective as any domain-specific language. We illustrate our ideas 
using the example of sparse Jacobian matrix computation. We implement an automatic 
differentiation method that takes advantage of sparse system structures and is 
straightforward to parallelize in a distributed memory setting. Furthermore, we show that the 
computational cost scales linearly with the size of the system.}


\maketitle

\section{Introduction}  \label{sec:introduction}

Differential-algebraic equations (DAEs) are ubiquitous in systems engineering 
problems, especially in design applications 
\cite{crawley2008contrasting,wetter2009modelica,wu2010fast,karimi2004modeling,lee2002power}. 
Mathematical models in this area 
are typically heterogeneous and very sparse. Obtaining the sparse Jacobian for such 
problems is critical for successful solving or preconditioning strategies.

Additional challenges arise from the need to effectively manage the complexity of system 
engineering models. The model equations are typically not in a single central place, but
assembled from component model equations loaded from multiple dynamic libraries.
Furthermore, the component model structure can (and often does) change at runtime. 
For example, when designing a heat exchanger one may want to keep inlet and outlet temperatures 
constant at operating conditions and optimize the heat exchanger geometry
parameters. In the transient simulations, however, the heat exchanger geometry is fixed and 
temperatures are system variables. The variable and parameter designation is selected by
the designer as needed at runtime. The ability to reuse the same model for 
different types of analyses is required to reduce the cost of the computation deployment as
well as the cost of component model verification and validation.

To address these requirements, domain-specific modeling languages based on symbolic code 
manipulations such as gPROMS \cite{oh1996} and Modelica \cite{olsson2012} have been 
introduced. Tools built around these languages \cite{dymola,wolfram,maple,gproms,matlab} 
allow engineers to work in a more interactive design environment 
where they can make modifications of their models at runtime. These tools go beyond Jacobian 
generation and perform a number of transformations on the mathematical model, such as causalization,
tearing, and index reduction to make subsequent simulations more efficient (see e.g.~\cite{cellier2006} and references therein). Under the hood, the model 
encoded in the domain-specific language is processed symbolically and code compatible 
with the numerical solver is generated and compiled on the fly. Then, such a hard-wired precompiled 
model is simulated and the result is returned to the user. This approach was pioneered in 
compiler automatic differentiation tools such as ADIC \cite{bischof1997adic} and 
OpenAD/F \cite{utke2008}.

Symbolic manipulations and 
compiling automatically generated code on the fly allow one to reuse models coded in a 
domain-specific language for different types of analyses using 
different numerical solvers. The downside is that the model needs to
be regenerated and recompiled every time the model structure is modified. Scaling up this 
approach to more complex problems is another 
challenge as the symbolic preprocessing of model equations may become a bottleneck. 
In such a framework 
one needs to support two different parallelization schemes -- one for the symbolic 
manipulations of the model equations and another one for solving these equations 
numerically. Symbolic transformations are generally nontrivial to parallelize. The more 
features the domain-specific language offers, the more complex the symbolic
processing algorithms become and so does their parallel implementation.
At the time of this writing, we are not aware of distributed memory parallel schemes for the
symbolic processing of mathematical equations. 

Modern object-oriented languages, such as C++, which support operator overloading, 
template specialization, type traits, and other advanced features allow one to create numerical 
models that can be reconfigured at runtime. In this paper we argue that the same functionality
provided by symbolic 
preprocessing of the model equations can be implemented by creating custom data types and
appropriate libraries in mainstream object-oriented languages.
Recently, solver frameworks that use abstract data types were proposed \cite{baker2012}. 
Those frameworks do not require specific data types to be used, but only specify 
elementary algebra that the data types have to support. By designing models and solvers 
to use abstract data types, one can reuse the same models and solvers for multiple analysis types
such as forward simulations, optimization, sensitivity analysis, or embedded uncertainty 
quantification \cite{constantine2014efficient}. Switching between these may be accomplished 
simply by changing (or reconfiguration of) the data type. Furthermore, abstract data types
can be used to compute automatically the system 
connectivity graph which could then be utilized, for instance, to partition the system
into smaller subsystems, perform index reduction for differential-algebraic equations, 
implement tearing algorithms, and many other calculations.

We illustrate this abstract elementary algebra approach by developing a method for 
sparse automatic Jacobian generation, which 
is superior to other available methods when applied to systems engineering problems. 
We show that our method allows for model reconfiguration at runtime 
and overall better code reuse in scientific applications. Moreover,
we show that our approach enables the automatic generation of the dependency graph 
of a system. The method is straightforward to parallelize in a distributed memory environment.

A similar approach is used in Sacado \cite{phipps2012}, an automatic differentiation
package which is part of the Trilinos library \cite{heroux2005}. However, Sacado does 
not support sparse derivatives -- it allocates memory for derivatives 
with respect to all system variables. 
For many partial differential equation (PDE) models this is not a significant limitation since the
sparsity pattern typically consists of locally dense cells and dense derivatives can be computed 
for each cell separately.
Automatic differentiation also can be incorporated directly in the mesh generation 
(e.g.~as in \cite{schoberl2014}), therefore using sparsity information from the mesh itself. 
On the other hand, components in system engineering applications are often large with locally sparse structures that can be exploited by using sparse automatic differentiation.

The automatic differentiation library ADOL-C \cite{walther2012} provides sparse
capability and has drivers for computing Jacobian and Hessian matrices. However, ADOL-C  does 
not define abstract elementary algebra, but uses operator overloading only locally to compute 
derivatives. Such approach requires ADOL-C specific code in the model, and is not easily extensible. 
Furthermore, ADOL-C does not support distributed memory parallelization.

The paper is organized as follows: In Section~\ref{sec:ad} we provide the mathematical
formulation of the elementary algebra we use for sparse automatic differentiation. We
describe the prototype C++ implementation in Section \ref{sec:prototype}. Preliminary 
benchmarking results showing linear scaling of the computational cost 
are presented in Section \ref{sec:benchmarking}. Future 
research directions are discussed in Section~\ref{sec:conclusion}.

\section{Mathematical Formulation of Sparse Automatic Differentiation}
\label{sec:ad}

\subsection{Problem Description}

In addition to the system of nonlinear equations or differential-algebraic equations, given by
\begin{equation}
    f(x; p) = 0 \quad \text{or} \quad F(t, x, \dot{x}; p) = 0, \label{eq:residual}
\end{equation}
respectively, most numerical libraries for the solution and optimization of such systems require the user to provide also the Jacobian and the sparsity pattern of the system. Here, $ x $ is the vector of variables and $ p $ the vector of parameters. The Jacobian $ J $ is then defined as the matrix
\begin{equation}
    J = \pd{f}{x}
    \quad \text{or} \quad
    J = \alpha \frac{\partial F}{\partial \dot x} + \frac{\partial F}{\partial x},
    \label{eq:jacobian}
\end{equation}
where the parameter $ \alpha $ is related to the numerical integration scheme used for the simulation and provided by the solver. There are several different ways to compute the Jacobian:
\begin{itemize}[leftmargin=1em]
\item \textit{Compute and implement analytical Jacobian manually}. While this will lead to the fastest numerical computations, it is often not feasible or extremely cumbersome to compute Jacobians manually for large systems.
\item \textit{Compute Jacobian numerically}. The derivatives can be computed using finite difference approximations. This approach is relatively easy to implement, but it is in general neither accurate nor efficient.
\item \textit{Use compiler automatic differentiation}. This approach requires equation syntax parsing capabilities. For large systems, parsing the equations can be quite time consuming. Parallelizing these methods could be quite challenging, as well.
\item \textit{Use operator overloading automatic differentiation\index{automatic differentiation}}. In this approach, all derivatives are computed automatically at the same time when the model equations are evaluated. This approach requires minimal involvement from the component model developer.
\end{itemize}
Another requirement is often to be able to reuse model equations in cases where some of the parameters $ p_i $ are set as variables and some of the variables $ x_i $ are ``fixed''  to constant values. Each of those cases would require different Jacobians. This complicates implementing the first and third approach for such systems significantly.

For most engineering problems, the governing equations are sparsely coupled. Thus, only a small fraction of the entries of the Jacobian will be nonzero. There are a number of different algorithms for solving linear systems that take advantage of the system sparsity to speed up computations \cite{li05,davis2010}. To use those algorithms, however, one also needs to find the \textit{sparsity pattern}.

\begin{example} \label{ex:Lorenz}
Assume we are trying to find a steady state solution for a Lorenz system \cite{lorenz1963}. The residual equations \eqref{eq:residual} can be written as
\begin{equation} \label{eq:lorenz}
    \begin{split}
        \sigma (y - x)   &= 0, \\
        x (\rho - z) - y &= 0, \\
        x y - \beta z    &= 0,
    \end{split}
\end{equation}
where $ x $, $ y $, and $ z $ are variables and $ \sigma $, $ \rho $, and $ \beta $ are constant parameters. The Jacobian \eqref{eq:jacobian} is then given by
\begin{equation} \label{eq:jacobian_lorenz}
 J =
\begin{bmatrix}
    -\sigma  &  \sigma &    0    \\
    \rho - z &  -1     &  -x     \\
    y        &  x      &  -\beta
\end{bmatrix}.
\end{equation}
Here, $J_{13} \equiv 0$, while all the other entries have nonzero values, generally. The sparsity pattern for the Lorenz system is then
\begin{equation} \label{eq:sparsity_pattern}
\begin{bmatrix}
    \bullet & \bullet &         \\
    \bullet & \bullet & \bullet \\
    \bullet & \bullet & \bullet
\end{bmatrix}.
\end{equation}
This tells the solver it does not need to allocate memory for $ J_{13} $ and perform computations with it. \exampleSymbol
\end{example}

Developing an efficient way for computing the sparsity pattern of the Jacobian is the key enabling technology for solving large-scale nonlinear systems, ordinary differential equations, differential-algebraic equations, and optimization problems. Since all of the information required to compute the Jacobian is contained within the model equations, the computation of the Jacobian and its sparsity pattern can be fully automated.

\subsection{Automatic Structure Analysis}

For better clarity, let us first discuss the sparsity pattern generation alone. The sparsity pattern is required by the numerical solver at the initialization stage to allocate objects required for sparse linear algebra algorithms. During computations, the sparsity pattern (i.e.\ connectivity structure) is used by the linear solver to identify structurally nonzero elements of the Jacobian that enter the computation.

The system connectivity information can be used for a number of other analyses such as index reduction for differential-algebraic equations (DAEs), partitioning, model causalization, tearing, numerical diagnostics, and many others. These are, however, beyond the scope of this paper and will be only briefly described in Subsection~\ref{subsec:Further Applications}.

The approach we propose is to compute the dependencies of the residual equations on the fly along with the
residual values. To do that, we
define a mathematical object $\mathcal{Y}$ which is a set containing a real number 
$y$ and set $\mathbb{D}$, which contains integer labels of all dependencies of $y$. 
Labels are independent variable identifiers; typically, they are offset values in the 
solution vector as returned by the solver. We denote this object as
\begin{equation}
    \mathcal{Y} = \{y, \mathbb{D}_y\}. \label{eq:dependency_tracking}
\end{equation}
For any independent variable $x$, the corresponding dependency tracking object is
\begin{equation}
    \mathcal{X} = \{x, \{n_x \} \},
\end{equation}
that is, each independent variable has only trivial self-dependency. Algebraic 
operations on $\mathbb{Y}$, the set of all $\mathcal{Y}$, are defined as follows:
\begin{itemize}[leftmargin=1em]
\item For any $C \in \mathbb{R}$ and $\mathcal{Y} \in \mathbb{Y}$ and algebraic operation 
$*$, it is
\begin{equation}
    C * \mathcal{Y} = \{ C * y, \, \mathbb{D}_y \}.
\end{equation} 
\item For any two $\mathcal{Y}, \, \mathcal{Z} \in \mathbb{Y}$ and mathematical operation 
$*$, it is
\begin{equation}
    \mathcal{Y} * \mathcal{Z} = \{ y * z, \, \mathbb{D}_y \cup \mathbb{D}_z \}.
\end{equation}
\item For any function $h(y)$ defined on $\mathbb{R}$, there is a corresponding function 
$h(\mathcal{Y})$ defined on $\mathbb{Y}$ such that
\begin{equation}
    h(\mathcal{Y}) = \{ h(y), \, \mathbb{D}_y \},
\end{equation}
where $\mathbb{D}_y$ is the set of dependencies of $\mathcal{Y}$.
\end{itemize}
Comparisons between elements of $\mathbb{Y}$ are performed with respect to values 
only, disregarding dependencies. For example:
\begin{equation}
    \mathcal{Y}_1 > \mathcal{Y}_2 \,\, \Leftrightarrow \,\, y_1 > y_2.
\end{equation}
If we define residual equations on $\mathbb{Y}$, rather than $\mathbb{R}$, the 
residual computation will give us both, the residual value and the sparsity pattern.

\begin{example}
Let us consider again the Lorenz system introduced in Example~\ref{ex:Lorenz}. The first residual is computed as
\begin{equation}
    \begin{split} \label{eq:lorenz1_dep}
        \mathcal{F}_1 &= \sigma \left( \{x, \{n_x \}\} - \{y, \{n_y \}\} \right) \\
                      &= \sigma \{ x - y, \{n_x \} \cup \{n_y \} \}              \\
                      &= \{ \sigma(x - y), \{n_x, n_y \} \},
    \end{split}
\end{equation}
and similarly we get
\begin{align}
    \mathcal{F}_2 &= \{x (\rho - z) - y, \{n_x, n_y, n_z \} \}, \label{eq:lorenz2_dep} \\
    \mathcal{F}_3 &= \{x y - \beta z,     \{n_x, n_y, n_z \} \}. \label{eq:lorenz3_dep}
\end{align}
From the dependencies in (\ref{eq:lorenz1_dep}--\ref{eq:lorenz3_dep}), one can obtain the sparsity pattern \eqref{eq:sparsity_pattern} by setting $n_x = 1$, $n_y = 2$, and $n_z = 3$. Note that this approach for obtaining the sparsity pattern is independent of how equations are written. If we, for example, write the third residual in \eqref{eq:lorenz} as  
\begin{equation}
    f_3 = u - \beta z, \label{eq:lorenz3_u}
\end{equation}
where $ u = x y $, then the residual evaluation using dependency tracking variables gives us
\begin{equation}
    \begin{split}
        \mathcal{F}_3 &= \{ u, \mathbb{D}_u \} - \beta \{ z, \{n_z\} \}  \\
                      &= \{ xy, \{n_x \} \cup \{n_y \} \} - \{ \beta z, \{n_z\} \} \\
                      &= \{ xy - \beta z, \, \{n_x, n_y\} \cup \{n_z\} \} \\
                      &= \{ xy - \beta z, \, \{n_x, n_y, n_z \} \},
    \end{split}
\end{equation}
which is the same as \eqref{eq:lorenz3_dep}. \exampleSymbol
\end{example}

This property is particularly convenient when coding residual equations because it allows reordering computations and using as many intermediate dependent variables as necessary. Note that all derivatives in the equations are uniquely defined in terms of derivatives with respect to independent variables, only.

\subsection{Automatic Differentiation}

To perform sparse automatic differentiation we make a small extension to the object we used for the sparsity pattern computation. We define 
$\tilde{\mathbb{Y}}$ as a set of all 
\begin{equation}
    \tilde{\mathcal{Y}} = \{y, \, \{(n, \partial_n y) : n \in \mathbb{D}_y\}\},
    \label{eq:ad}
\end{equation}
where $y$ and $\mathbb{D}_y$ are the same as in \eqref{eq:dependency_tracking}.
Essentially, we mapped to each dependency the value of the partial derivative with
respect to that dependency. For independent variables 
\begin{equation}
    \tilde{\mathcal{X}} = \{x, \, \{(n_x, 1)\} \}.
\end{equation}
Algebraic operations on $\tilde{\mathbb{Y}}$ are defined in a similar 
fashion as in the dependency tracking case:
\begin{itemize}[leftmargin=1em]
\item For any $C \in \mathbb{R}$, $\tilde{\mathcal{Y}} \in \tilde{\mathbb{Y}}$ 
and algebraic operation $*$ defined on $\tilde{\mathbb{Y}}$, it is
\begin{equation} \label{eq:const_times_var}
    C * \tilde{\mathcal{Y}} = \{ C * y, \, \{(n, \partial_n (C * y)) : n \in \mathbb{D}_y\}\}.
\end{equation}
\item For any two $\tilde{\mathcal{Y}}, \tilde{\mathcal{Z}} \in \tilde{\mathbb{Y}}$ 
and algebraic operation $*$ defined on $\tilde{\mathbb{Y}}$, it is
\begin{equation} \label{eq:var_times_var}
    \tilde{\mathcal{Y}} * \tilde{\mathcal{Z}}
     = \{ y*z, \, \{(n, \partial_n (y * z)) : n \in \mathbb{D}_y \cup \mathbb{D}_z \} \}.
\end{equation}
\item For any function $h(x)$ defined on $\mathbb{R}$, there is a corresponding function 
$h(\tilde{\mathcal{Y}})$ defined on $\tilde{\mathbb{Y}}$ such that
\begin{equation}
    h(\tilde{\mathcal{Y}}) = \{ h(y), \, \{ (n, h'(y) \partial_n y) : n \in \mathbb{D}_y \}\},
\end{equation}
where $\mathbb{D}_y$ is the set of dependencies of $\tilde{\mathcal{Y}}$.
\end{itemize}
Comparisons between elements of $\tilde{\mathcal{Y}}$ are defined in the same way as 
for the dependency tracking data type. 

\begin{example}
Let us compute the Jacobian for residual \eqref{eq:lorenz3_u}.
Using the automatic differentiation data type defined in \eqref{eq:ad}, and the 
algebra defined for it, this computation is carried out as
\begin{equation*}
    \begin{split}
        \mathcal{F}_3 &= \{ u, \{ (n_x,\partial_x u), (n_y, \partial_y u) \} \} 
                       - \beta \{ z, \{(n_z, 1) \} \}                            \\
                      &= \{ xy, \{(n_x, y), (n_y, x)\} \} 
                       - \{ \beta z, \{ (n_z, \beta) \} \}                       \\
                      &= \{ xy - \beta z, \, \{(n_x, y), (n_y, x), (n_z, -\beta) \} \}.
    \end{split}
\end{equation*}
The derivatives in $\mathcal{F}_3$ make up the third row of the Jacobian
\eqref{eq:jacobian_lorenz}, when $n_x = 1$, $n_y = 2$, and $n_z = 3$. \exampleSymbol
\end{example}

\subsection{Further Applications}
\label{subsec:Further Applications}

\begin{figure}[htb]
    \centering
    \begin{minipage}{0.3\textwidth}
        \centering
        {\scriptsize{a)}} \\[1mm]
        \includegraphics[width=\textwidth]{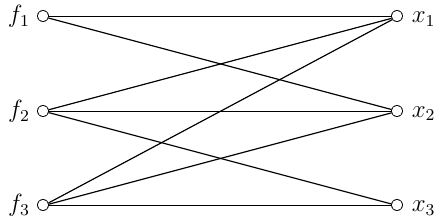}
    \end{minipage}
    \hspace*{1cm}
    \begin{minipage}{0.346\textwidth}
        \centering
        {\scriptsize{b)}} \\[1mm]
        \includegraphics[width=\textwidth]{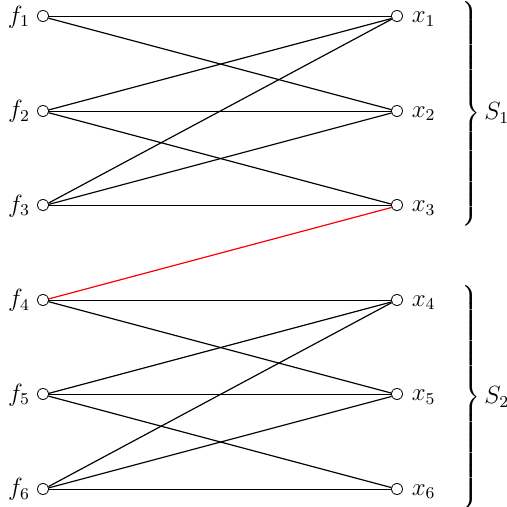}
    \end{minipage}
    \caption{a) Dependency graph of the Lorenz system. b) Partitioning of two coupled Lorenz systems using the dependency graph. Only the edge marked in red needs to be shared by the two cores.}
    \label{fig:DependencyGraph}
\end{figure}

Analogously, the structure of a given system can be represented as a bipartite graph, where the bipartite sets of vertices are the equations and variables, respectively. Equation $ f_i $ is then by definition connected to variable $ x_j $ if and only if $ J_{ij} \not \equiv 0 $. The dependency graph of the Lorenz system \eqref{eq:lorenz} is shown in Figure~\ref{fig:DependencyGraph}a. These dependency graphs are typically used for symbolic transformations such as causalization and tearing of model equations. The causalization selects the order in which equations are solved, so that system can be solved gradually, piece by piece. Tearing is typically used in conjunction with causalization to break dependency loops. These methods are described in detail in~\cite{cellier2006}, for example. With our data type, the dependency graph is automatically generated during the model evaluation and these methods can be implemented without source code transformation tools.

The dependency graph could also be used to decompose a high-dimensional system into smaller subsystems. With the aid of graph partitioning tools, it would be possible to break the system into smaller clusters that are strongly connected internally, but only weakly connected to the other clusters. In this way, assigning different clusters or subsystems to different processing cores, the communication between cores -- which is often a bottleneck during the parallel simulation of such systems -- could be minimized without domain knowledge simply by using a data type that tracks the dependencies. An illustration of this approach is shown in Figure~\ref{fig:DependencyGraph}b, where two coupled Lorenz systems are split into subsystems $ S_1 $ and $ S_2 $. Here, only variable $ x_3 $ is required by both subsystems and needs to be shared by the two cores if the system is solved in parallel. Note that subsystem $ S_1 $ can be evaluated independently whereas $ S_2 $ depends on $ S_1 $. This can also be regarded as a block causalization of the system.

\section{Prototype Implementation} \label{sec:prototype}

Dependency tracking objects like \eqref{eq:dependency_tracking} or \eqref{eq:ad} can be seamlessly implemented in 
any programming language that supports operator overloading\index{operator overloading}. 
We created a preliminary implementation in C++ mainly for prototyping and testing 
purposes. Here, we outline the details of this implementation.

We create a class \texttt{Variable} that stores the double precision value and the dependency map 
related to that value. The class overloads all operators defined for the double 
data type. The prototype implementation is structured like this:
\begin{lstlisting}
class Variable
{
public:
    Variable();
    explicit Variable(double value);
    Variable(double value, size_t variableID);
    Variable(const Variable& v);
    ~Variable();
    
    // =
    Variable& operator=(const double& rhs);
    Variable& operator=(const Variable& rhs);
    
    // +=
    Variable& operator+=(const double& rhs);
    Variable& operator+=(const Variable& rhs);
    
    // *=
    Variable& operator*=(const double& rhs);
    Variable& operator*=(const Variable& rhs);
    
    // ...
    
    typedef std::map<size_t, double> DependencyMap;    
    
private:
    double value_;
    size_t variableID_;
    bool isFixed_;
    
    mutable DependencyMap* dependencies_;    
};
\end{lstlisting}
This class has several constructors. The constructor that creates Variable from double 
data type is made explicit to prevent a possible loss of derivatives in 
accidental implicit data conversions. In addition to the value and the
identifier of the variable, the object has a boolean flag \verb|isFixed_|. This flag is used when the
designation of the object needs to change from a variable to a constant parameter. 
The dependency map in this implementation is just a standard map between the variable 
identifiers and the values of the derivatives with respect to those variables. Only independent
(state) variables have assigned identifiers. The identifiers typically correspond to global vector 
indices. Dependent or temporary variables 
would obtain their dependency map directly or indirectly from state variables. For 
example $x$, $y$, and $z$ in equation \eqref{eq:lorenz3_u} would have identifiers set to $1$, $2$, and $3$, respectively, but $u$ would not have an identifier.

All arithmetic operators are overloaded for \verb|Variable| data type to implement 
\eqref{eq:var_times_var} and combinations of \verb|Variable| and \verb|double| 
types to implement \eqref{eq:const_times_var}.
Arithmetic operators are implemented in terms of compound assignment operators. For 
example, here is how \texttt{*=} operator overloading is implemented for the Variable class:
\begin{lstlisting}
Variable& Variable::operator*=(const Variable& rhs)
{
    // derivation by parts of @ *this
    scaleDependencies(rhs.value_);

    // compute partial derivatives of rhs and add them to *this
    for(auto& p : *(rhs.dependencies_))
        (*dependencies_)[p->first] += (p->second * value_);

    // compute value of *this
    value_ *= rhs.value_;

    return *this;
}
\end{lstlisting}
Here, we compute the derivative $(lhs \cdot rhs)' = (lhs)' \cdot rhs + lhs \cdot (rhs)'$, where $lhs$ is pointed to by \texttt{this}. The derivatives of $lhs$ are first scaled by the value
of $rhs$ to obtain the first term in the expression for the derivative of the product. 
Then each $rhs$ derivative is multiplied by the value of $lhs$ and added to corresponding
derivatives of $lhs$. If the corresponding derivative of $lhs$ does not exist,
a new entry is added to the dependency map.
The value is computed at the end, because it is used in the
derivative of the product expression.

The multiplication operator is then implemented as a non-member operator
\begin{lstlisting}
const Variable operator*(const Variable& lhs, const Variable& rhs)
{
    return Variable(lhs) *= rhs;
}
\end{lstlisting}
The copy constructor with \texttt{lhs} as the argument is used to create the variable to be 
returned and then the \texttt{*=} operator is used to multiply that variable by \texttt{rhs}.

In addition to overloading operators, all mathematical functions from the standard C++ library 
operating on double data type have to be overloaded, as well. Take for example the
sine function whose derivative is $(\sin x)' = \cos x \cdot x'$:
\begin{lstlisting}
namespace std
{
    inline Variable sin(const Variable& x)
    {                                         
        double val = sin(x.getValue());                         
        double der = sin_derivative(x.getValue());               
        Variable res(x);            // copy derivatives of x
        res.setValue(val);          // set function value f(x)
        res.scaleDependencies(der); // compute derivatives of f(x)
        return res;
    }
}
\end{lstlisting}
In the namespace \texttt{std} we define an inline function that takes a
Variable data type as the input. We use the copy constructor to retain 
all derivatives of $x$, and then we multiply them by $\cos x$ per 
chain rule. The value of the sine is computed using the sine function from the
standard C++ library. The function that computes the derivative of the sine
is defined in the same namespace as the Variable type:
\begin{lstlisting}
inline double sin_derivative(double x)
{
    return std::cos(x);
}
\end{lstlisting}
The same approach is used for other functions. 

\begin{example}
A simple use of the Variable class is shown in the following code:
\begin{lstlisting}
template <typename T>
void residualFunction(vector<T>& f,
                      const vector<T>& x,
                      const vector<T>& p)
{
    const T y = x[0]*x[1];
    
    f[0] = p[0]*(x[1] - x[0]);        // sigma*(y - x)
    f[1] = x[0]*(p[1] - x[2]) - x[1]; // x*(rho - z) - y
    f[2] = y - p[2]*x[2];             // x*y - beta*z
}

int main()
{
    const size_t n = 3;
    vector<Variable> x(n), p(n), f(n);
    
    // initialize independent variables
    x[0] = 8.0; x[1] = 20.0; x[2] = 2.0/3.0;    
    
    // set constant parameter values
    p[0] = 10.0; p[1] = 8.0/3.0; p[2] = 28.0;
    
    // decide x, y, and z are variables... 
    for (size_t i = 0; i < n; ++i)
        x[i].setVariableNumber(i);
    
    // ...and sigma, rho, and beta are constant parameters    
    for (size_t i = 0; i < n; ++i)
        p[i].setFixed(true);
    
    residualFunction(f, x, p);
    printIncidenceMatrix(f);
    printJacobian(f);
}
\end{lstlisting}  
The function \texttt{residualFunction} computes the residual for the steady state solution of the Lorenz system. The dependent variable \texttt{y} is not really needed in this example other than to illustrate how the derivative calculation is propagated through variables. It is important to note that the Lorenz model in this implementation does not depend on a specific data type. Furthermore, the model does not assume what are independent variables and what are system parameters. This is determined outside the model. In this example, elements of the vector \texttt{x} are set to be independent variables and elements of the vector \texttt{p} constant parameters in the main function. This is consistent with the problem defined in \eqref{eq:lorenz} where we look for a steady state solution given parameters $ \sigma $, $ \rho $ and $ \beta $.
One residual evaluation with the Variable data type also computes the sparsity pattern and the Jacobian. The function \texttt{printIncidenceMatrix} outputs the sparsity pattern:
\begin{lstlisting}
    A = [1 1 0;
         1 1 1;
         1 1 1];
\end{lstlisting}
The function \texttt{printJacobian} outputs:
\begin{lstlisting}[mathescape]
    J = [-10  10   0;
           2  -1  -8;
          20   8 -28]; |\Comment{\exampleSymbol}|
\end{lstlisting}
\end{example}

Postprocessing of this output may help, for instance, detect (structurally) singular or 
ill-conditioned Jacobians. For simplicity, we omit the implementation of the two output functions.
Note that in this example the parameters \texttt{p} could be declared as \texttt{double} instead 
of \texttt{Variable}, thus getting rid of a small overhead when calling \texttt{p}. In that case, 
however, we would need to modify \texttt{residualFunction} code if we wanted to change 
parameter and variable designation as in the next example.

\begin{example}
The \texttt{Variable} data type allows us to change parameter and variable designations at 
runtime, so we can designate elements of the vector \texttt{p} as system variables and elements 
of the vector \texttt{x} as constant system parameters. All one has to do is to set:
\begin{lstlisting}
    // decide x, y, and z are constant parameters... 
    for (size_t i = 0; i < n; ++i)
        x[i].setFixed(true);
        
    // ...and sigma, rho, and beta are variables    
    for (size_t i = 0; i < n; ++i)
        p[i].setVariableNumber(i);
\end{lstlisting}
In this problem we look for parameters $ \sigma $, $ \rho $, and $ \beta $ such that a fixed point solution of~\eqref{eq:lorenz} is at $ x = 8 $, $ y = 20 $, and $ z = 2/3 $. The sparsity pattern and Jacobian of the system in that case are obtained as
\begin{lstlisting}
    A = [1 0 0;
         0 1 0;
         0 0 1];
\end{lstlisting}
and
\begin{lstlisting}
    J = [12  0  0;
          0  8  0;
          0  0 -0.667];
\end{lstlisting}
respectively. This change can be made at runtime without the need to recompile the system model. Any other selection of system parameters and variables can be made in the same way. The vectors \texttt{x} and \texttt{p} are used merely to denote \textit{nominal} system variables and parameters.
\exampleSymbol
\end{example}

Typically, one would run the residual evaluation with dependency tracking during the solver initialization phase
to get the sparsity pattern, and then run the residual evaluation with derivative calculation every time the
Jacobian is required during solver iterations. Note that the residual vector
in this case is in fact an implementation 
of a compressed row sparse matrix. The only difference is that each row, 
in addition to matrix elements, also holds a corresponding residual vector 
element.

\section{Preliminary Benchmarking Results} \label{sec:benchmarking}

Our approach provides model developers with an intuitive interface, 
simplifies porting of legacy codes, allows for model equations to be 
modified at runtime and is straightforward to parallelize. The trade-off 
is that allocating variables in local scope is expensive and adds significant 
new overhead to the computation. For our approach to be feasible, the 
overhead has to be manageable and scale well with the size of the system.
To provide a preliminary 
assessment of how the computational cost scales with the size of the model 
(i.e.\ number of equations), we perform several numerical experiments. 
For the purpose of this assessment, we treat all variables (including residuals) 
as local scope variables and reallocate derivatives each time they are computed.

\begin{figure}[htb]
    \centering
    \includegraphics[width=0.7\textwidth]{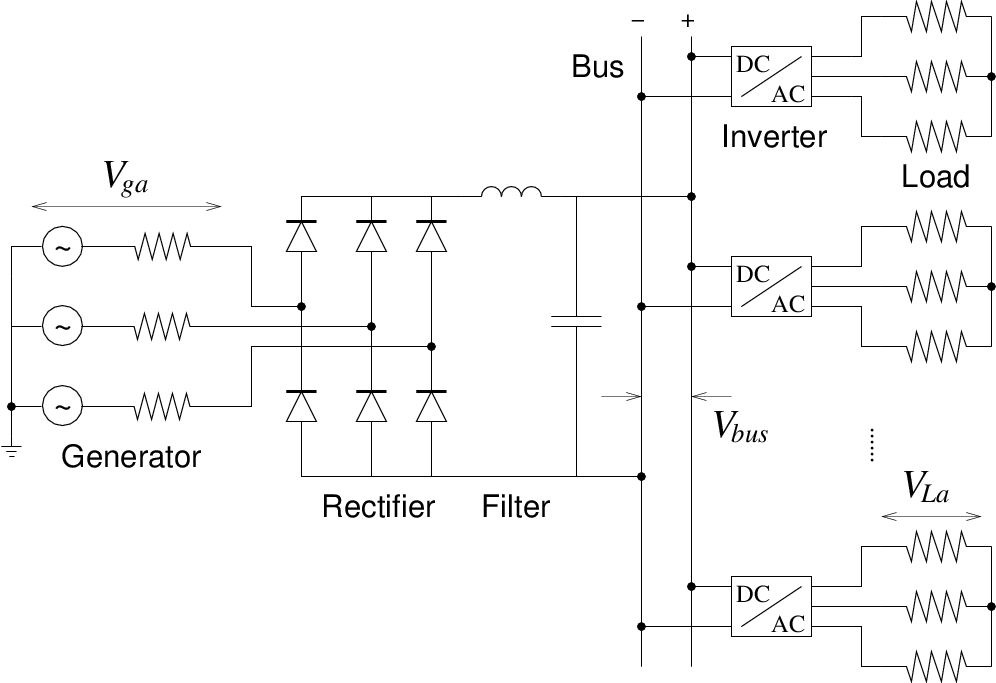} 
    \caption{Benchmarking test case: A simple electrical grid model.}
    \label{fig:grid}
\end{figure}

As a benchmark problem, we select a simple microgrid model 
as shown in Figure~\ref{fig:grid}. A single alternating current (AC) generator is connected
to a rectifier that converts power to direct current (DC) and supplies it to a DC bus. 
Several passive AC loads are connected to the DC bus, each through a separate 
inverter that converts DC power from the bus to 60\unit{Hz} AC power required 
by the load. The size of this system can be easily scaled up by simply 
adding more loads to the bus. The system parameters are set so that the
simulation results are ``self-validating'' -- the voltage at each load 
is the same as the voltage produced by the generator: 100\unit{V}, 60\unit{Hz} 
sinusoidal. A more detailed description of the test case is given in Appendix 
\ref{sec:RectifierInverter}.

The electrical grid model is cast in form of differential-algebraic equations. 
We simulate the first 0.1\unit{s} of the grid operation using the Rythmos package from
the Trilinos library. We use an implicit variable-order variable-stepsize backward 
differentiation method to solve the equations \cite{hindmarsh2005}.
The solutions of the resulting systems of nonlinear equations are obtained using a sparse direct method.
The method requires the residual equations and the Jacobian of the system to be provided. We measure the overall computation
time and model evaluation time. In our case, a model evaluation is a residual 
evaluation using the Variable data type. The Jacobian is evaluated automatically
together with the residual at each model evaluation call per design of the Variable 
class. 

\begin{figure}[htb]
    \centering
    \includegraphics[width=0.7\textwidth]{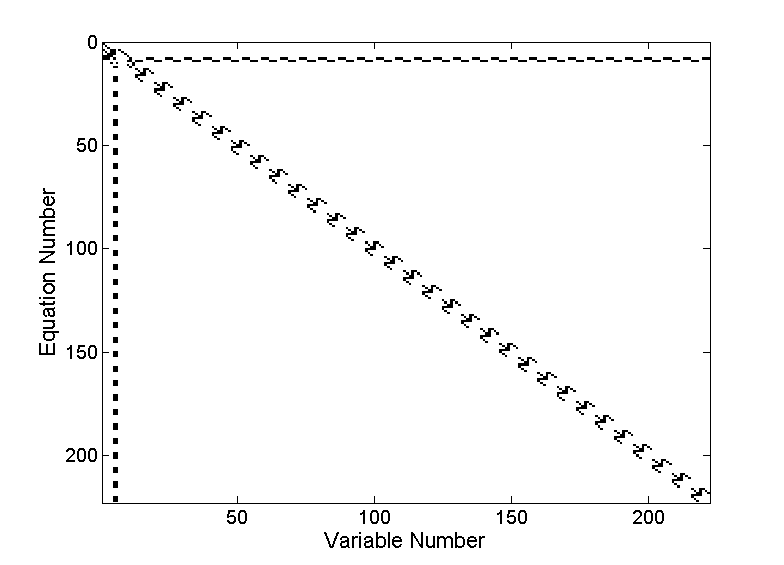} 
    \caption{Sparsity pattern for the grid model with 30 loads. Only 
    $1.8\,\%$ of Jacobian elements are structurally nonzero. }
    \label{fig:sparsity_pattern}
\end{figure}

The unknown variables $ x $ in our case are typically node voltages. An example of the Jacobian sparsity pattern for the benchmarking test case is shown in Figure~\ref{fig:sparsity_pattern}.

\begin{figure}[htb]
    \centering
    \includegraphics[width=0.7\textwidth]{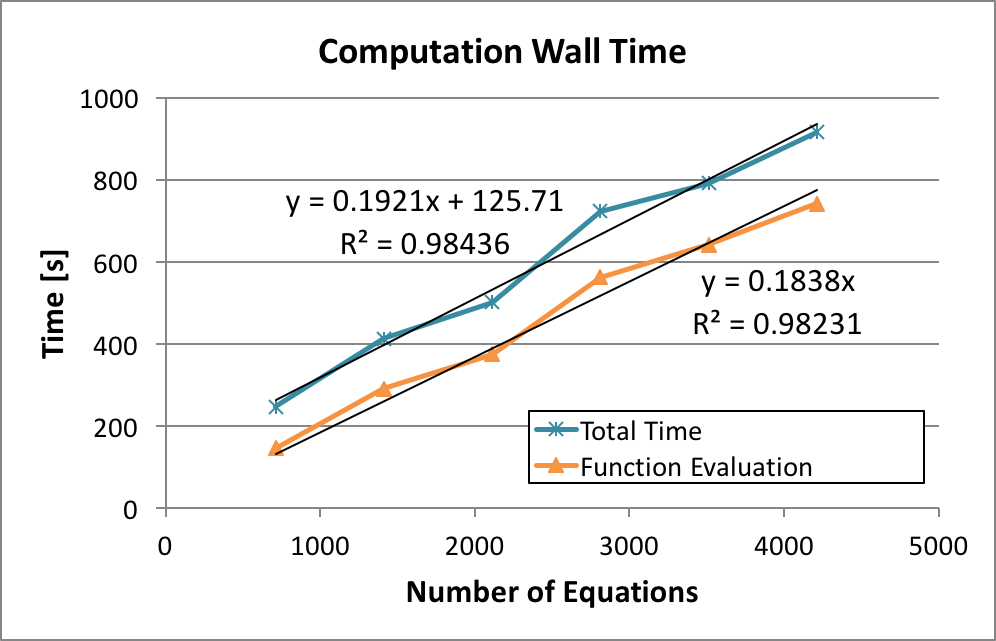} 
    \caption{Wall time for the overall computation and the model evaluation only.}
    \label{fig:serial}
\end{figure}

In the serial case, we ran simulations for grids with 100--600 loads. For these 
simulations it takes roughly 8,000 integrator steps and 50,000--60,000 function 
evaluations to complete regardless of the size of the system.
We find that the average computational time per call grows \textit{linearly} with 
the size of the system as shown in Figure \ref{fig:serial}. 
Data points for the computational cost of the function evaluation fit particularly 
well to the linear model. 

\begin{figure}[htb]
\centering
    \includegraphics[width=0.7\textwidth]{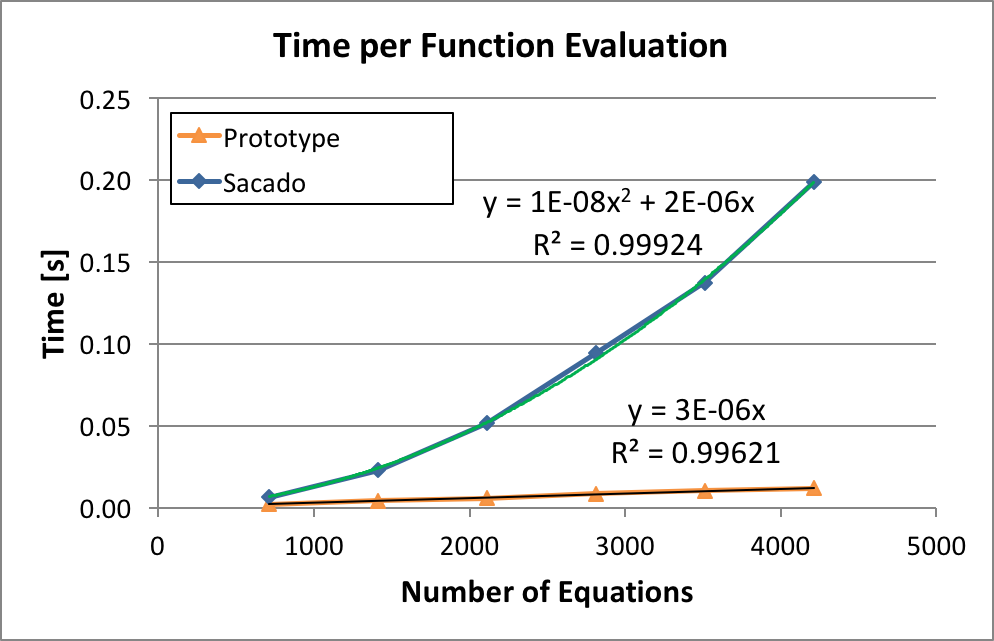} 
    \caption{Function evaluation cost when using sparse and dense automatic differentiation.} \label{fig:sacado}
\end{figure}

We compare the average cost of the function evaluation for our sparse automatic 
differentiation prototype with dense automatic differentiation using the
Sacado package from the Trilinos library. The prototype uses the class \verb|map| from the standard
C++ library ($O(\log n)$ cost), whereas Sacado uses a dense vector ($O(1)$ cost)
to store and access derivatives. We use our method to compute sparsity pattern 
information and provide it to Sacado, so that only structurally nonzero derivatives are computed. 
For systems of this size, it was expected that the dense algorithm would 
outperform the sparse automatic differentiation. Both algorithms evaluate 
the same derivatives and the dense approach has faster access to derivatives. 
The only downside of the dense approach is that it has to allocate larger 
chunks of memory to store derivatives. Yet, our results suggest that 
memory management alone may cause the computational cost to grow quadratically 
with the size of the system when using dense automatic differentiation 
(Figure \ref{fig:sacado}). 

\begin{figure}[htb]
    \centering
    \includegraphics[width=0.7\textwidth]{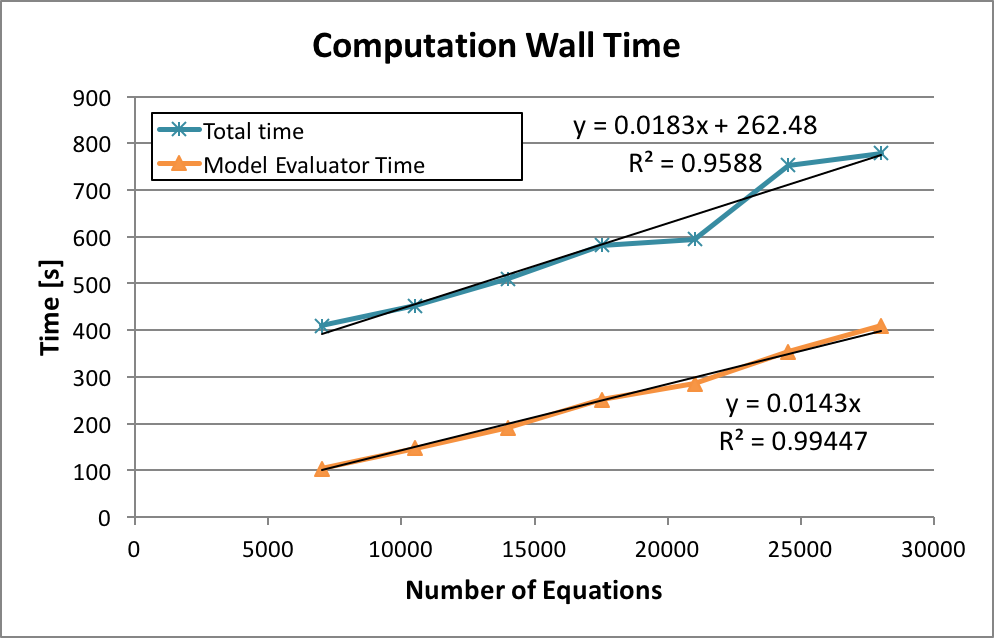} 
    \caption{Computational time per call for integration step and function evaluation.}
    \label{fig:parallel}
\end{figure}

Parallelizing simulations that use sparse automatic differentiation 
to compute the Jacobian is fairly straightforward in the MPI framework when using our approach. For testing purposes, we implemented a simple parallelization scheme where the
generator and the rectifier are simulated on one node and the simulations of 
inverters and loads are evenly distributed over the remaining nodes. We 
show here results of an MPI simulation on 16 CPU cores (4 nodes). 
The number of loads in the system was varied from 1,000--4,000 (roughly 10,000--30,000 equations). The results show again linear scaling as the size of the system increases 
as shown in Figure \ref{fig:parallel}. 
Sacado eventually uses up all the memory on the 
compute node in this case and simulation crashes.

\section{Conclusion} \label{sec:conclusion}

Our analysis and benchmarking results suggest that using the proposed abstract
elementary algebra approach for sparse automatic differentiation is 
a promising direction. The linear scaling of the computational costs and
the ease of parallelization indicate that this approach is particularly
suitable for massively parallel computations. 

The prototype implementation leaves room for code optimization. The 
residual vector implemented in terms of the Variable data type is \textit{de facto}
a compressed row sparse Jacobian with the extension that each row
is associated with the corresponding residual value. 
At this time we have not implemented an option to switch off Jacobian computation
when the solver does not need it.
Intermediate variables, such as
$u$ in \eqref{eq:lorenz3_u}, can be understood as sparse matrix rows, as well. 
However, they are not part of the Jacobian, they are used just to complete the 
chain rule. Typically, intermediate variables are used by system modelers 
for convenience to write model equations in a more compact form. In the current 
implementation, derivatives of residual functions are reallocated at every 
function evaluation based on the dependency tracking mechanism. Since this information
does not change during solver iterations, the dependency structure of the system 
could be precomputed once and then reused at subsequent solver iterations.
This could be done easily for residual vectors, which are typically passed by 
reference to models. Intermediate variables are typically local variables used by
the modeler to simplify the equations, so they may have to be reallocated at every 
iteration anyway. The code optimization and more thorough performance testing will be 
done in a subsequent work.

When simulating our test cases, we had to copy the residual values from the 
vector of variables to the Epetra vector and the Jacobian derivatives to a
compressed-row sparse matrix provided by Epetra, per requirements of the Rythmos solver.
This added a small additional overhead to the computation. The full power
of the proposed approach could be demonstrated with numerical solvers that 
do not require specific data formats, but instead provide abstract interfaces 
to all linear and elementary algebra operations. While such solvers are 
still not part of the mainstream, a lot of activities have been done in that 
direction as for example in Tpetra project \cite{baker2012}.

Using abstract elementary algebra has potential applications way beyond 
automatic differentiation. As we have shown in this paper, it could be 
used for reconfiguring models at runtime; constant parameters could be 
changed into variables and vice versa. Abstract data types also 
could be used for diagnostics, for example to identify structurally singular 
Jacobians. Furthermore, this approach could be used for preprocessing 
model equations for index reduction of differential-algebraic equations or 
tearing algorithms for system decomposition. 

Abstract elementary algebra can also help reuse existing code. The same model code
can be reused for local sensitivity analysis or embedded uncertainty 
quantification, simply by using different template parameters. The same 
holds true for solvers that provide abstract interfaces for elementary 
algebra. More reuse streamlines code verification and improves 
development efficiency, which are critical for any large scale 
computation. All of these will be pursued in subsequent work.

\appendix

\section{Test Case Description} \label{sec:RectifierInverter}

Electrical grids  are a common motif in power systems. An example of such a grid is shown in Figure~\ref{fig:grid}.
Since mathematical modeling methods for power 
systems are not commonly known outside the electrical engineering 
community, we provide here a brief overview of the governing 
equations used in our test case. For a detailed description of 
the component models, we refer the reader to, for example, \cite{erickson2001}
and references therein. The equations are derived and the system 
is composed using the modified nodal analysis approach \cite{riaza2008}.

We assume that the generator shown on the left side in Figure \ref{fig:grid} 
is an ideal 3-phase generator with residual equations for the generator terminals $a$, $b$, and $c$ given by:
\begin{align}
    0 &= v_{ga} - V_0 \sin (\omega t),  \label{eq:first} \\
    0 &= v_{gb} - V_0 \sin (\omega t + 2\pi / 3),        \\
    0 &= v_{gc} - V_0 \sin (\omega t + 4\pi / 3).
\end{align}

\begin{figure}[htb]
    \centering
    \includegraphics[width=0.35\textwidth]{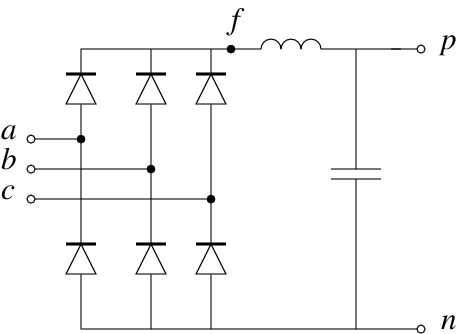}
    \caption{Schematic of the rectifier model.}
    \label{fig:rectifier}
\end{figure}

The generator is connected to a rectifier, which converts the 3-phase 
AC power to DC power. Rectifier and filter schematics are shown in
Figure \ref{fig:rectifier}. 
Kirchhoff's current law for the rectifier can be cast in terms of 
residual equations as
\begin{align}
    0 &= i_{ga} - I_D(v_{ga} - v_f) + I_D(v_n - v_{ga}), \\
    0 &= i_{gb} - I_D(v_{gb} - v_f) + I_D(v_n - v_{gb}), \\
    0 &= i_{gc} - I_D(v_{gc} - v_f) + I_D(v_n - v_{gc}).
\end{align}
Here, $i_{ga}$, $i_{gb}$, and $i_{gc}$ are generator phase currents 
entering the rectifier nodes $a$, $b$, and $c$;
$v_f$ is the voltage at node $f$ (rectifier's positive terminal,
connected to the filter) and $v_n$ is the node voltage at 
rectifier's negative terminal. 

The current through diodes $I_D(v)$ in the rectifier is modeled as
\begin{equation*}
    I_D(v) = I_s \left[\exp\left(\frac{v}{nkT}\right) - 1 \right].
\end{equation*} 
In our simulations, we chose the temperature $T=300\,$K, ideality factor $n=2$, 
and saturation current to be $I_s = 18.8\,$nA. Boltzmann constant 
$k \approx 1.38 \times 10^{-23}$ J/K. The voltage $v$ in the
diode current function is the difference between node voltages at the anode
and cathode terminals of the diode.

The residual equations for the filter can be written as
\begin{align}
    0 &= I_D(v_{ga} - v_f) + I_D(v_{gb} - v_f) + I_D(v_{gc} - v_f) - i_L, \\
    0 &= \phi_L - L i_L,                                                  \\  
    0 &= \dot \phi_L - (v_f - v_p),                                       \\
    0 &= q_C - C(v_p - v_n),
\end{align}
where $i_L$ and $\phi_L$ are inductor's current and flux, respectively, 
$L$ is the inductance, $q_C$ is charge on the capacitor, $C$ is the 
capacitance, and $v_p$ is the voltage at the positive terminal of the DC bus.

Kirchhoff's current law gives the following equations for the DC bus (Figure~\ref{fig:grid}):
\begin{align}
    \begin{split}
    0 = {}& -\dot q_C + i_L \\
        & + \sum_k G_k \left(  d^{(k)}_a(t) v^{(k)}_a + d^{(k)}_b(t) v^{(k)}_b + d^{(k)}_c(t) v^{(k)}_c  \right),
    \end{split} \\
    \begin{split}
    0 = {}& I_D(v_n - v_{ga}) + I_D(v_n - v_{gb}) + I_D(v_n - v_{gc}) \\
        &  + \sum_k \left( i^{(k)}_a + i^{(k)}_b + i^{(k)}_c \right),
    \end{split}
\end{align}
where $v^{(k)}_\alpha$ and $i^{(k)}_\alpha$ are $\alpha$-phase of the 
voltage across and current through the load $k$ ($\alpha = a,b,c$).
The functions $d^{(k)}_\alpha(t)$ are inverter modulation signals that
describe 3-phase AC waveforms at the inverter outlet. We choose the modulation 
signals to be the same for all inverters and produce the sinusoidal output:
\begin{align*}
    d^{(k)}_a (t) &= m \sin(\omega t),            \\
    d^{(k)}_b (t) &= m \sin(\omega t + 2\pi / 3), \\
    d^{(k)}_c (t) &= m \sin(\omega t + 4\pi / 3).
\end{align*}
Here, we set $\omega$ to be the same as the frequency of the 
generator. Furthermore, we set
\begin{equation*}
    m = \frac{2\pi}{3\sqrt{3}},
\end{equation*}
so that the voltage amplitude at each load is the same as the generator 
voltage amplitude. This choice was made merely for model verification 
convenience (again, interested readers are referred to \cite{erickson2001} for more details). 

\begin{figure}[htb]
    \centering
    \includegraphics[width=0.35\textwidth]{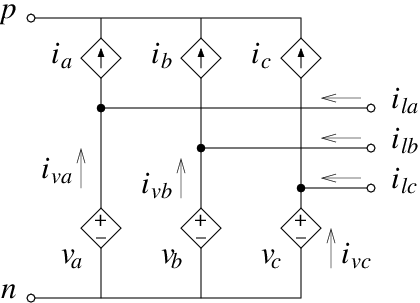}
    \caption{Schematic of the inverter model with averaged pulse width modulation.}
    \label{fig:inverter_averaged}
\end{figure}

The inverter model we used is an averaged model without details of 
pulse width modulation. The equivalent circuit model of such an
inverter model is shown in Figure \ref{fig:inverter_averaged}. 
The model consists of three ideal current-controlled current sources 
and three ideal voltage-controlled voltage sources (one for each phase). 
The current sources are controlled by the load currents as
\begin{equation*}
    i_\alpha = d_\alpha (t) i_{l \alpha}, \quad \alpha = a,b,c,
\end{equation*}
where $i_{l \alpha}$ is the load current (Figure \ref{fig:inverter_averaged}).
Voltage sources are controlled by the DC bus voltage as
\begin{equation*}
    v_\alpha = \frac{1}{2} d^{(k)}_\alpha(t)(v_p - v_n), \quad \alpha = a,b,c.
\end{equation*}
Here, $v_p$ and $v_n$ are the node voltages on the positive and negative terminals of 
the DC bus, respectively. By using Kirchhoff's current law, the equations for 
the inverter are obtained as
\begin{align}
    0 &= i^{(k)}_{va} - G_k (v^{(k)}_a - v^{(k)}_0) 
           + d^{(k)}_a(t) G_k (v^{(k)}_a - v^{(k)}_0), \\
    0 &= i^{(k)}_{vb} - G_k (v^{(k)}_b - v^{(k)}_0) 
           + d^{(k)}_b(t) G_k (v^{(k)}_b - v^{(k)}_0), \\
    0 &= i^{(k)}_{vc} - G_k (v^{(k)}_c - v^{(k)}_0) 
           + d^{(k)}_c(t) G_k (v^{(k)}_c - v^{(k)}_0), \\
    0 &= G_k (v^{(k)}_a - v^{(k)}_0) + G_k (v^{(k)}_b - v^{(k)}_0) 
       + G_k (v^{(k)}_c - v^{(k)}_0).
\end{align}
Using Kirchhoff's voltage law, we obtain equations for voltages at load 
terminals $v_{l \alpha}$:
\begin{align}
    0 &= v^{(k)}_{la} - v_n -\frac{1 + d^{(k)}_a(t)}{2}(v_p - v_n), \\
    0 &= v^{(k)}_{lb} - v_n -\frac{1 + d^{(k)}_b(t)}{2}(v_p - v_n), \\
    0 &=  v^{(k)}_{lc} - v_n -\frac{1 + d^{(k)}_c(t)}{2}(v_p - v_n). \label{eq:last}
\end{align}

The entire system is described by residual equations 
(\ref{eq:first}--\ref{eq:last}). There are $12 + 7N$ system variables, 
where $N$ is the number of AC loads connected to the bus. System variables 
are the generator node voltages $v_{ga}$, $v_{gb}$, and $v_{gc}$; the generator 
currents $i_{ga}$, $i_{gb}$, and $i_{gc}$; the rectifier node voltage $v_f$;
the positive and negative voltages $v_p$ and $v_n$ of the DC bus; the internal filter
variables -- the inductor current $i_L$, the inductor flux $\phi_L$, and the capacitor
charge $q_C$; currents through ideal voltage sources in the inverter
model $i^{(k)}_{va}$, $i^{(k)}_{vb}$, and $i^{(k)}_{vc}$; and load node 
voltages $v^{(k)}_{la}$, $v^{(k)}_{lb}$, $v^{(k)}_{lc}$, and $v^{(k)}_{l0}$.
Index $k=1,\ldots,N$ denotes a load. In our tests, we set parameters to 
the following values: Generator frequency $\omega = 2 \pi 60 \,$rad/s, load 
conductances are all equal and set to $G=0.01\,$S, capacitance in the filter 
is $C=0.1\,$mF, and inductance in the filter is $L=20\,$mH. 

\section*{Acknowledgments}
The authors would like to thank Eric Phipps of Sandia National Laboratories 
for his helpful suggestions and productive discussions. Warm thanks go to 
Teems Lovett of United Technologies Research Center
for a number of helpful discussions and proofreading of the final draft. 

\bibliographystyle{plain}
\bibliography{PelesKlus2017}

\end{document}